# Single-nanowire, low-bandgap hot carrier solar cells with tunable open-circuit voltage

Steven Limpert[1*], Adam Burke[1], I-Ju Chen[1], Nicklas Anttu[1], Sebastian Lehmann[1], Sofia Fahlvik[1], Stephen Bremner[2], Gavin Conibeer[2], Claes Thelander[1], Mats-Erik Pistol[1] and Heiner Linke[1*]

[1]NanoLund and Solid State Physics, Lund University, Box 118, 22100 Lund, Sweden

[2]School of Photovoltaic and Renewable Energy Engineering, University of New South Wales, 2052 Sydney, Australia

[*]email: steven.limpert@ftf.lth.se, heiner.linke@ftf.lth.se

**Abstract**

Compared to traditional pn-junction photovoltaics, hot carrier solar cells offer potentially higher efficiency by extracting work from the kinetic energy of photogenerated "hot carriers" before they cool to the lattice temperature. Hot carrier solar cells have been demonstrated in high-bandgap ferroelectric insulators and GaAs/AlGaAs heterostructures, but so far not in low-bandgap materials, where the potential efficiency gain is highest. Recently, a high open-circuit voltage was demonstrated in an illuminated wurtzite InAs nanowire with a low bandgap of 0.39 eV, and was interpreted in terms of a photothermoelectric effect. Here, we point out that this device is a hot carrier solar cell and discuss its performance in those terms. In the demonstrated devices, InP heterostructures are used as energy filters in order to thermoelectrically harvest the energy of hot electrons photogenerated in InAs absorber segments. The obtained photovoltage depends on the heterostructure design of the energy filter and is therefore tunable. By using a high-resistance, thermionic barrier an open-circuit voltage is obtained that is in excess of the Shockley-Queisser limit. These results provide generalizable insight into how to realize high voltage hot carrier solar cells in low-bandgap materials, and therefore are a step towards the demonstration of higher efficiency hot carrier solar cells.

**Introduction**

When a semiconductor of bandgap $E_G$ absorbs a photon, the portion of the photon energy exceeding $E_G$ becomes kinetic energy of the photogenerated electron and hole. In pn-junction solar cells, this excess kinetic energy is transferred as waste heat to the lattice by electron-phonon interaction and cannot be converted to electrical potential energy [1,2]. To avoid this energy loss, and to potentially increase the maximum power conversion efficiency to as much as 85% [3], it has been suggested to extract work from hot carriers before they cool to the lattice temperature [2–6]. Specifically, it was predicted that a thermoelectric contribution to device voltage would be present when a photoinduced temperature gradient is present across a carrier-energy filtering heterostructure [7,8]. In this way, hot-carrier solar cells can recover a portion of the ~400 mV voltage loss attributable to the cooling of carriers from 6000 K to 300 K [2], and thus allow larger voltages than those achievable in conventional single-junction solar cells made of the same materials.

Work towards the realization of hot carrier solar cells has proceeded in many directions. Transport of photogenerated carriers through Si quantum dots in $SiO_2$ has been investigated [9,10]. Ultra-fast, hot electron collection has been demonstrated in bandgap engineered GaAs/AlGaAs heterostructures [11,12] and hot carrier transport across an InP/PbSe interface has been studied [13]. Hot carriers have been spectroscopically observed and predicted to result in solar cell efficiency enhancement in GaAsP/InGaAs quantum wells [14] and hot carrier photocurrent has been observed in a GaAs/InGaAs quantum well solar cell [15]. Ferroelectric insulators have been demonstrated to exhibit above bandgap photovoltages [16] and barium titanate, $BaTiO_3$, has been shown to exhibit power conversion efficiencies in excess of the Shockley Queisser limit [17] due to hot carriers and the bulk photovoltaic effect [18,19].

All of the above demonstrations of extraction of photogenerated, hot carriers have been performed in materials with a relatively large bandgap (i.e. $E_G > 1$ eV). However, the maximum power conversion efficiency achievable with a hot carrier solar cell depends upon the bandgap of the material, and the theoretically achievable efficiency in hot carrier solar cells is the highest in low-bandgap materials (i.e. $E_G < 0.5$ eV) [3,4]. Recently, we reported single-nanowire, photothermoelectric devices that produced bipolar currents under illumination by different wavelengths of light [20]. Here, we point out that these devices are in fact, low-bandgap hot carrier solar cells as they were made of wurtzite (WZ) InAs, which has a room temperature bandgap of only 0.39 eV [21,22]. In this work, we expand upon the discussion of these devices and show that they are hot carrier solar cells. We do this by comparing their measured current voltage (I-V) curves to the Shockley-Queisser [23] detailed balance limit for an ideal pn-junction solar cell composed of the same absorbing material and showing that the open-circuit voltage of the highest resistance single-barrier device exceeds this limit. Then, we discuss the energy conversion process that allows achievement of this limit-breaking photovoltage. Next, we demonstrate that photovoltage tunability through heterostructure engineering is a characteristic of the presented low-bandgap hot carrier solar cells by showing that when we increase energy filter transmissivity, we increase device conductivity and we decrease the achievable open-circuit voltage. Finally, we discuss topics for future work.

**Methods: Device Design and Fabrication**

The devices in this study are based on single nanowires with either a single- or double-barrier heterostructure acting as an energy filter (Fig. 1). The basic principle for the generation of photocurrent and photovoltage in these hot carrier solar cells is illustrated in Fig. 1b,e and relies on: (i) energy filters that separate photogenerated hot carriers (Figs. 1c and 1d), and (ii) absorption hot spots forming near the filters to give rise to photogenerated carriers in their vicinity (Figs. 1b and 1e). This localized increase in carrier concentration is possible because light absorption in a nanowire is not homogenous, but concentrated in hot spots corresponding to maxima of electromagnetic wave modes [20]. Electron-hole pairs are photogenerated predominantly in these hot spots and there establish a local non-equilibrium carrier temperature that can be much higher than the lattice temperature [20,24,25]. When an absorption hot spot is located within a hot-carrier diffusion length of a few hundred nanometers from an energy filter, hot electrons can diffuse across the filter before cooling. This charge movement results in a measurable photocurrent from which electrical power can be extracted (Fig. 1b) and the separation of photogenerated electrons and holes leads to the formation of an open-circuit voltage (Fig. 1e).

Nanowires are ideal for hot carrier solar cells for several reasons. First, their optical properties are highly tunable [26]: the concentration and confinement of light inside the nanowire (i.e. photonic effects) can be combined with the electromagnetic generation of surface-confined, oscillating electron plasmas at metal-dielectric interfaces (i.e. plasmonic effects) to control the position of spatially well-defined photon-absorption hot spots within the nanowire. This enables the ideal, nearby positioning of energy filters for fast carrier separation and work extraction (Fig. 1b, e). Second, because of radial strain relaxation, nanowires are more amenable to bandgap engineering than planar devices [27,28]. This enables heterostructures of materials of desirable bandgaps and band offsets to be selected and fabricated with atomic precision and with low defect densities. Third, likely because of reduced electron-phonon interaction in nanowires, the temperature of photogenerated carriers can be much higher than that of the lattice [20,25]. Finally, a single-nanowire device setup enables the use of a backgate (Fig. 1f) to tune the carrier concentration during experiments [29]. This enables us to experimentally access different conductivity regimes within a single device.

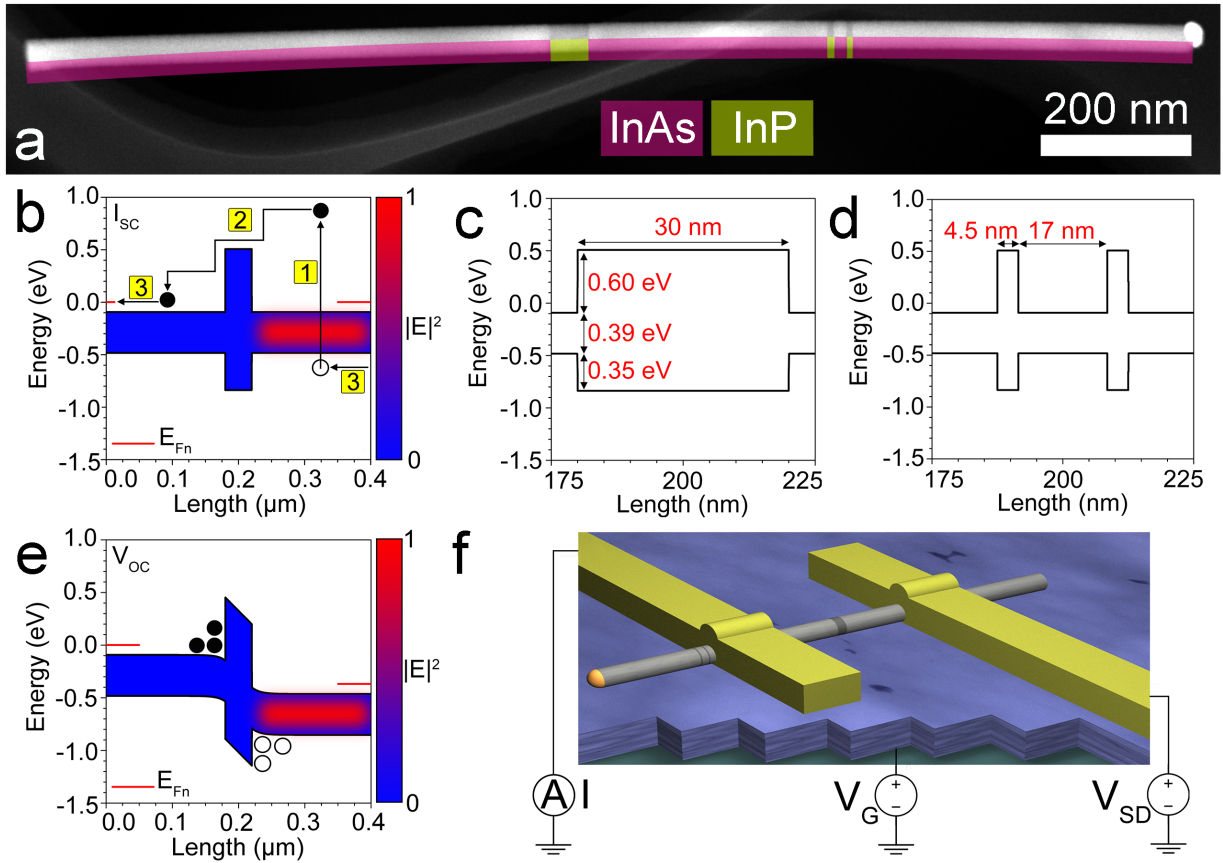

**Figure 1 | Single nanowire hot carrier solar cell. A**, Scanning transmission electron high angle annular dark field micrograph of an InAs/InP heterostructure nanowire with single- and double-barrier InP segments that each can act as energy filter. In any given device, only one of the segments is used as an energy filter; the other one is not contacted. InAs and InP segments are false colored (InAs – pink, InP – yellow) as a guide to the eye. **b**, Band diagram of a single-barrier device under short-circuit current conditions. The red area indicates the location of a hot spot of photon absorption and carrier generation. Steps 1-3 indicate the process of current generation: (1) photogeneration of an electron-hole pair; (2) diffusion of a hot electron across the barrier, followed by thermalization; (3) the electron leaves to the left and drives a current through the circuit, filling the photogenerated hole from the right. The electron quasi-Fermi levels, $E_{Fn}$, at the contacts are indicated by red lines. **c,d** Band diagrams under short-circuit conditions and geometry of the heterostructures used in this work. **e**, Band diagram of a single-barrier device under open-circuit voltage conditions at a bias of 0.37 V. **f**, 3D illustration of a single barrier hot carrier solar cell with electrical measurement circuit. Spacing between the inner edges of the contacts is 400 nm.

In the proof-of-principle demonstration of Ref. 19, we used wurtzite (WZ) InAs as the absorber material because of its small bandgap, $E_G$ = 0.39 eV [21,22], corresponding to light with bandgap wavelength $\lambda_G$ = 3180 nm, allowing absorption of a broad spectrum of light. Furthermore, InAs exhibits high electron-hole asymmetries of effective mass and mobility, enabling photogeneration of high-energy, fast-diffusing electrons and low-energy, slow-diffusing holes, thereby assisting in electron collection across the energy-filter and charge separation. As the barrier/energy-filter material, we used InP ($E_G$ = 1.34 eV, $\lambda_G$ = 925 nm) [30]. We further defined two types of InAs/InP heterostructures (Fig. 1a), namely (i) single, thermionic barriers because they are predicted to produce the highest thermoelectric power [31,32] (Fig. 1c) and (ii), double-barriers - which have been previously used in hot carrier solar cell experiments [11,12] - because of the energy filtering effect [33] of resonant tunneling structures (Fig. 1d).

InAs/InP nanowire heterostructures with atomically sharp interfaces were grown using chemical beam epitaxy (CBE) (Fig. 1a). Nanowires were transferred to an $SiO_2$ substrate equipped with a backgate, and we electrically contacted individual nanowires by electron beam lithography (Fig. 1f). Contacts were fabricated with a 400 nm inner separation, ensuring that hot carriers would only need to travel a maximum of about 200 nm to be collected across the heterostructure before they cooled – a much shorter distance than an estimated hot-carrier diffusion length in InAs (see Supporting Information for more information). The InAs material was naturally n-type and no pn-junction was present within the nanowires. Both types of energy filters used were grown into the same nanowires (Fig. 1a), and contacts were placed around the structure of interest in different devices (Fig. 1f). For clarity, in the following sections of this paper, devices in which contacts were placed around a double-barrier quantum dot will be referred to as double-barrier devices and devices in which contacts were placed around a single, thermionic barrier will be referred to as single-barrier devices.

Devices were electrically characterized in vacuum in a variable-temperature ($T$ = 6 K – 300 K) probe station with optical fiber access. DC electrical measurements were made using the measurement circuit shown in Fig. 1f using a Yokagawa 7651 DC source, a Stanford Research Systems SR570 current preamplifier, a Hewlett Packard 34401A multimeter and a Keithley 2636B SourceMeter. For photovoltaic characterization we used light generated by a supercontinuum laser and selected by a monochromator resulting in a Gaussian spectrum with a center wavelength of 720 nm and a full-width at half-maximum of 140 nm. Integration of the spectrum's spectral irradiance results in a computed irradiance of 17.6 kW/m$^2$ and integration of the spectrum's spectral photon flux results in a computed total above-bandgap photon flux of 6.77×10$^{22}$ photons/m$^2$ (see Supporting Information for method details).

**Results and Discussion**

Dark and illuminated current voltage (I-V) curves of the single-barrier device show that it was fabricated properly and that it produces electrical power when illuminated (Figure 2). The dark current-voltage curve of the single-barrier device is symmetric and exponentially increasing under both forward and reverse bias (Figure 2a). This is the characteristic current-voltage shape for thermionic emission over a barrier and confirms that the device does not contain a pn-junction. The figures of merit of the illuminated single-barrier device (Figure 2b) are as follows: short-circuit current, $I_{SC}$ = -13.3 ± 0.2 pA, open-circuit voltage, $V_{OC}$ = 368 mV ± 5 mV, and fill-factor, FF = 27.5 ± 0.4 %.

To place these results into context, we computed the dark and illuminated current voltage curves of an ideal pn-junction diode made of WZ InAs using the Shockley-Queisser detailed balance method (see Supporting Information for details). The calculated figures of merit of an ideal pn-junction solar cell made of WZ InAs that has the same projected area and surface area as the presented nanowire device and that is illuminated by the experimental spectrum are as follows: $I_{SC}$ = -165.6 pA, $V_{OC}$ = 251 mV, and FF = 68.7%.

Comparison between the $V_{OC}$ = 368 mV measured for the illuminated single-barrier device and that of an ideal pn-junction ($V_{OC}$ = 251 mV) provides strong evidence that hot-carrier energy conversion is essential to the voltage generation in the presented device, and enhances the achievable voltage compared to a pn-junction made of the same contacted absorber material. Our interpretation is that in the presented device kinetic energy of hot photogenerated electrons is converted into voltage based upon a thermoelectric effect [7,8,20], extracting electrical power from the differential in carrier temperature across the thermionic barrier. Because of this mechanism, hot carrier solar cells are not bound by the Shockley Queisser detailed balance limit, which assumes isothermal energy conversion.

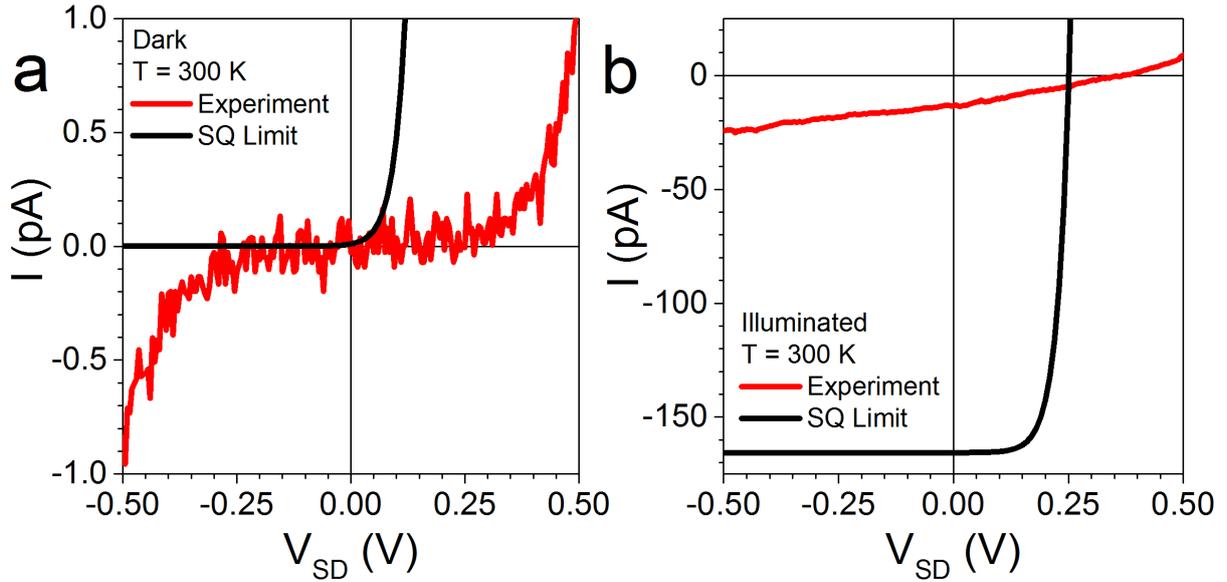

**Figure 2 | Dark and illuminated current voltage curves of an experimental single-barrier device and of an ideal pn-junction diode made of WZ InAs computed using the Shockley-Queisser detailed balance method. a**, Measured current voltage curve of the single-barrier device in the dark at room temperature (red curve). Calculated current voltage curve of an ideal pn-junction diode made of WZ InAs computed using the Shockley-Queisser detailed balance method in the dark at room temperature (black curve). **b**, Measured current voltage curve of the single-barrier device under illumination at room temperature (red curve). Calculated current voltage curve of an ideal pn-junction diode made of WZ InAs under the same illumination at room temperature computed using the Shockley-Queisser detailed balance method (black curve).

While the single-barrier device exceeds the $V_{OC}$ of an ideal pn-junction solar cell made of WZ InAs, the measured $I_{SC}$ = -13.3 ± 0.2 pA and FF = 27.5 ± 0.4 % are much lower than those for the corresponding ideal pn-junction ($I_{SC}$ = -165.6 pA, FF = 68.7%). This results in a lower power conversion efficiency of the single-barrier device compared to an ideal pn-junction solar cell made of WZ InAs. There are three possible reasons for this smaller $I_{SC}$: first, the nanowire does not absorb all of the light that is incident upon its projected area as its diameter is too small to support guided modes at the illumination wavelength. Second, not all of the light that is absorbed is absorbed in the hot spot next to the energy filter. Third, some photogenerated electrons and holes are likely to recombine within the single-barrier device before they are separated across the energy filter. This could happen if (1) a hot electron cools before crossing the energy filter, (2) the cooled electron recombines with its hole before being recycled up to energies high enough to cross the energy filter or (3) the hot electron diffuses in the direction opposite to the energy filter and recombines without encountering the energy filter. The smaller FF in the single-barrier device compared to an ideal pn-junction solar cell made of WZ InAs is because the current voltage curve of the single-barrier device is linear in the power producing region. This is a characteristic of thermoelectric devices [34] and hot carrier solar cells based on the bulk photovoltaic effect [16–19].

While hot carrier solar cells are based on a thermoelectric effect, they offer opportunities for high-efficiency energy conversion that are different than those offered by traditional thermoelectrics. This is because the presence of hot carriers can lead to very large temperature differentials over very small distances and between different distributions of particles (e.g. electrons and phonons). In comparison to traditional thermoelectric devices - in which performance is limited by parasitic heat flow in the lattice [35] - the heat transfer to the lattice in a nanoscale hot carrier solar cell can be suppressed if hot carriers are extracted from the device before they cool to the lattice temperature, a process that can be enhanced if electron-phonon energy exchange is inhibited by phononic engineering. As discussed in Ref. [20], we estimate the differential in the electron (carrier) temperature in the presented devices to be 170 K across the single-barrier, a value that is consistent with measurements of the non-equilibrium carrier temperature sustained in photogenerated carrier populations generated in small diameter nanowires under steady-state illumination [25]. Such a large temperature gradient would not be sustainable in traditional thermoelectrics, where carriers and phonons generally are in local thermal equilibrium, and it significantly enhances the achievable thermoelectric energy conversion efficiency. Importantly, power optimization and efficiency limits of thermionic thermoelectrics have been studied [31,32,36,37] and it has been shown that maximum power can be achieved at $T_C$ = 300 K using a $k_x$ filter with a barrier height of $1.1 k_B T_H$ [31]. Given the estimated $T_H$ of 470 K, this corresponds to a barrier of 45 meV. In this optimal configuration, a thermoelectric efficiency limit at maximum power of ~38% of the Carnot efficiency is predicted [31], corresponding to ~14% efficiency for the given $T_C$ and $T_H$ – a result which is in agreement with the quantum bounds on thermoelectric power and efficiency [37].

How do hot carrier solar cells compare to pn-junction solar cells in terms of strategies to boost their open-circuit voltage? In pn-junction solar cells, increasing the open-circuit voltage requires the elimination of sources of non-radiative recombination in order to decrease bias-induced dark current and increase the 'turn on' voltage of the diode that comprises the solar cell. While reducing non-radiative recombination to increase short-circuit current is also important in hot carrier solar cells, of similar importance is engineering the energy-filtering, charge-separating heterostructure. To achieve a high open circuit voltage in a low-bandgap hot carrier solar cell, we find that it is necessary to have an energy filter that is highly resistive to low energy electrons and holes, while simultaneously highly transmissive to high energy electrons. An energy filter with these characteristics enables achievement of a large open-circuit voltage because (1) it prevents backflow leakage of cooled photogenerated electrons after they have transited the energy filter (2) it decreases the bias-induced dark current of the device and (3) it inhibits the movement of low-energy photogenerated holes, ensuring that ambipolar movement of photogenerated electron-hole pairs is avoided. These physics are embedded in the following expression which describes how in a planar, illuminated, power-producing device with a linear current voltage curve and a thickness, $d$, the open-circuit voltage is inversely proportional to the sum of the dark and illuminated conductivity [17], $\sigma_d$ and $\sigma_{pv}$, respectively:

$$V_{OC} = \frac{J_{SC}d}{\sigma_d + \sigma_{pv}} \qquad (1)$$

In short, in a hot carrier solar cell, the photovoltage can be tuned by engineering the conductivity of the energy-filtering, charge-separating heterostructure.

Indeed, in our experiments, we observe an increase in the device conductivity and a decrease in the achievable open-circuit voltage when we use a double-barrier quantum dot (Fig. 1c) instead of a single, thermionic barrier (Fig. 1d) as the heterostructure energy filter. The increased conductivity of the double-barrier device in comparison to the single-barrier device can be attributed to the many current-carrying, resonant energy levels that exist below the barrier height in the quantum dot between the double-barriers. These energy levels result in a room temperature, zero gate voltage conductance that is approximately four orders of magnitude greater than that of a single-barrier device (Fig. 3a). Because of this high conductivity, to observe power-producing photocurrents and photovoltages under illumination, it is necessary to cool the double-barrier devices to $T = 6$ K and to apply a back-gate voltage of $V_G = -20$ V to suppress dark conductivity. Even then, the high transmissivity of the double-barrier heterostructure results in high illuminated conductivity and therefore, a low maximum open-circuit voltage of only ~17 mV (Fig. 3b).

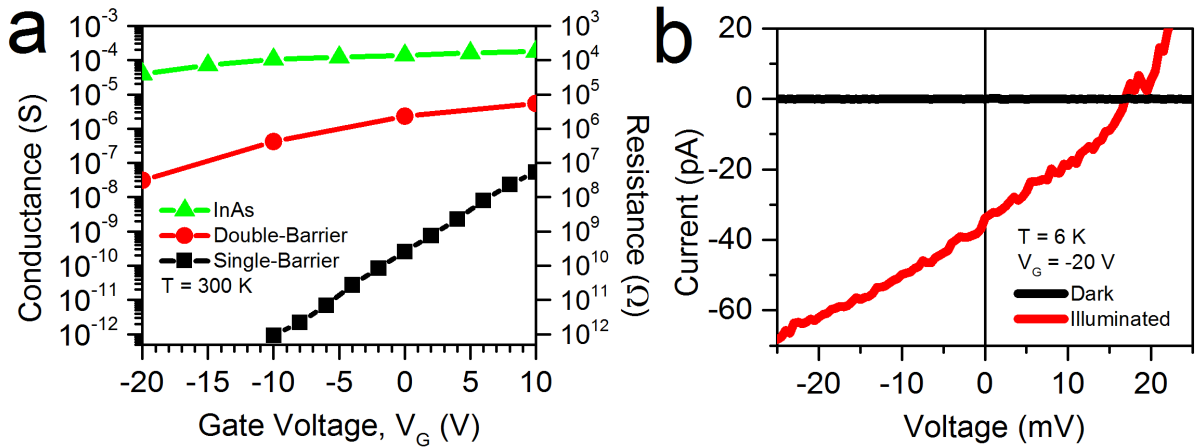

**Figure 3 | Dark conductance comparison and double-barrier current voltage curves. a,** Conductance (left axis) and resistance (right axis) of three devices: a pure InAs nanowire, a double-barrier device and a single-barrier device as a function of backgate voltage, $V_G$, at room temperature. **b**, Dark and illuminated current voltage curves of a double-barrier, quantum dot device at T = 6 K and $V_G$ = -20 V. In this regime, the nanowire is fully depleted and behaves as an insulator in the dark (black curve). Under illumination, the device is photoconductive and produces electrical power (red curve). In comparison to the single-barrier device, the double-barrier device exhibits a larger short-circuit current ($I_{sc}$ = -34 pA), but a smaller open-circuit voltage ($V_{oc}$ = 17 mV).

**Conclusion and Outlook**

We foresee several routes to increasing the short-circuit current and the fill-factor of the presented low bandgap nanowire hot carrier solar cells. To increase the short-circuit current, we anticipate that the following strategies may be useful: (i) increasing nanowire diameter to increase absorption, (ii) optically designing contacts to increase absorption and to concentrate absorption on one side of and nearby an energy filter, (iii) passivating the nanowire surface to increase electron mobility and lifetime (iv) optimizing the placement, height and width of the energy filter, (v) tailoring the nanowire diameter, crystal phase and heterostructures to minimize electron relaxation rates due to phonon scattering, and (vi) adding a hole contact to collect holes and reflect electrons. Additionally, we anticipate that these optimization techniques may be applied in the modeling-guided design of vertical nanowire arrays [38] in which photons are absorbed more strongly closer to the tips of the wires. Modeling suggests that it is possible to design nanowire diameter and array pitch such that a broadband absorption hot spot is present within the top 500 nm of the wires, where an energy filter could be placed within their hot-carrier diffusion length. Furthermore, additional concentration of longer wavelength light into this volume may be possible by use of plasmonic elements [39,40]. Finally, to increase the fill factor, we anticipate adding band bending into the device by doping or by local gating to induce nonlinearity in the illuminated current voltage curve.

Material choice will also play an important role in optimizing the devices described in this work. It is likely that by moving to absorbers with smaller bandgaps, higher carrier temperatures and efficiencies can be achieved as a larger fraction of photon energy will be converted into carrier kinetic energy. Moving to a wider bandgap barrier would likely enable larger open-circuit voltages by decreasing the thermionic dark current. However, maximum power has been predicted to be achieved with the estimated temperature difference at a barrier height of 45 meV [31], suggesting that a move to a narrower bandgap barrier material would be advantageous. In the end, to better understand the practical and the theoretical efficiency limits for these devices, and to determine the precise parameters of an ideal bandstructure, comprehensive optoelectronic and thermal device modelling will be required including self-consistent hydrodynamic simulations.

**Acknowledgements**

The authors thank Magnus Borgström, Gaute Otnes and Pyry Kivisaari for discussions, and acknowledge financial support by an Australian-American Fulbright Commission Climate Change and Clean Energy Scholarship and by a UNSW University International Postgraduate Award to S.L., by NanoLund, by Swedish Energy Agency (award no. 38331-1), by the Knut and Alice Wallenberg Foundation (project 2016.0089), by the Swedish Research Council (project no 2014-5490) by the Solander Program and by the Australian Government through the Australian Renewable Energy Agency (ARENA). Responsibility for the views, information, or advice expressed herein is not accepted by the Australian Government.

# Supporting Information

**Hot-carrier Diffusion Lengths**

In conventional photovoltaics, a figure of merit used to evaluate material quality and design devices is the diffusion length, $L$, the average distance that a photogenerated carrier travels before it recombines, namely

$$L = \sqrt{D\tau} \qquad (2)$$

where $D$ is the diffusivity of the particle and $\tau$ is the particle lifetime. If we replace $\tau$ by $\tau_{th}$, the time that it takes photogenerated carriers to thermalize amongst each other to form a thermal distribution, we can compute a thermalization diffusion length, $L_{th}$, an approximate average distance that a photogenerated carrier will travel during the process of carrier-carrier thermalization:

$$L_{th} = \sqrt{D\tau_{th}} \qquad (3)$$

Furthermore, if we replace $\tau$, with $\tau_{cool}$, the time that it takes photogenerated carriers to cool to the lattice temperature, we can compute a corresponding hot-carrier cooling diffusion length:

$$L_{cool} = \sqrt{D\tau_{cool}} \qquad (4)$$

These figures of merit provide useful information on the distances over which photogenerated hot-carrier transport can likely be experimentally observed.

To compute approximate values for these figures of merit for electrons in InAs, we can combine data on the diameter-dependent mobility of electrons in InAs nanowires ($\mu_n = 10^4 \text{ cm}^2/(\text{Vs})$) [1] with the Einstein relation ($D = \mu k_B T/e$) [2] and approximate thermalization by electron-electron interaction ($\tau_{th} = 1$ ps, the time scale for establishing a carrier temperature) and cooling times by electron-phonon interaction ($\tau_{cool} = 100$ ps, the time scale for carrier cooling to the lattice temperature) [3]. Assuming a temperature, $T$, of 300 K to establish a lower bound, we arrive at the following: $L_{th} = \sqrt{\mu k_B T \tau_{th}/e} \approx 160$ nm is the length scale on which photogenerated electrons in InAs establish an effective carrier temperature (which may be much higher than the lattice temperature) and $L_{cool} = \sqrt{\mu k_B T \tau_{cool}/e} \approx 1.6$ μm is the length scale on which electrons in InAs cool to the lattice temperature. In a hot-carrier solar cell, carrier separation must be achieved on a length scale less than $L_{cool}$.

**Illumination**

For photovoltaic characterization, we used a Fianium Femtopower 1060 Supercontinuum Source with emission from 500 nm to 1850 nm, a maximum power output of 8 W and a repetition rate of 82.5 MHz coupled into a Princeton Instruments SP2150 Double Monochromator. All presented measurements were performed using the monochromator's grating with 150 lines per millimeter, 800 nm blaze. Illumination spectra were measured with an Avantes Avaspec-3648-usb2 silicon CCD spectrometer and the power of narrowband slices of the source emission were measured with a Thorlabs power meter (item number: PM100D) and silicon and germanium photodiodes (item numbers: S120C and S122C).

The spectral irradiance, $F$, of the spectrum that was used to illuminate our devices and that was used as the input to our Shockley-Queisser detailed balance model was obtained in the following manner from the measured spectra and narrowband powers. The measured narrowband power, $P$ (Figure S1a), was linearly interpolated and divided by the product of the cross-sectional area of the optical fiber, $A = \pi \cdot (100 \text{ μm})^2$ to compute irradiance. In this calculation of irradiance, we neglected divergence of the beam. This is a valid assumption as the fiber tip was placed directly atop the sample during illuminated current voltage measurements. Therefore, there was negligible distance over which the beam could diverge. Then, to calculate spectral irradiance from irradiance, the irradiance was multiplied by 0.954 and divided by the narrowband bandwidth, $W$, of 60 nm, which was determined from a Gaussian fitting of the measured narrowband spectrums (Figure S1b). This calculation of spectral irradiance assumes an equal contribution to power from each wavelength within the narrowband, which is a valid assumption as the supercontinuum source emission power as a function of wavelength is flat, and is based on the fact that 95.4% of the area of a Gaussian curve is contained within two standard deviations on either side of the peak. This spectral irradiance was then multiplied by a Gaussian with amplitude of 1, center at 720 nm and standard deviation of 60 nm as determined from a fitting of the normalized broadband experimental spectrum spectrometer data (Fig. S1c), $G = exp\left(-\frac{1}{2}\left(\frac{x-720 \text{ nm}}{60 \text{ nm}}\right)^2\right)$. This multiplication of the broadband spectral irradiance by a narrower-band Gaussian captures the effect of the monochromator selecting a portion of the available spectrum. Thus, in the end, the spectral irradiance, $F$, is given by the expression

$$F = 0.954 \frac{PG}{AW} \qquad (5)$$

The computed spectral irradiance, $F$, of the spectrum used to illuminate our devices and used as the input to the Shockley Queisser detailed balance model is presented in Fig. S2. Its integration results in an irradiance of 17.6 kW/m². Dividing by photon energy and integrating gives a total photon flux of $6.77 \times 10^{22}$ photons/m².

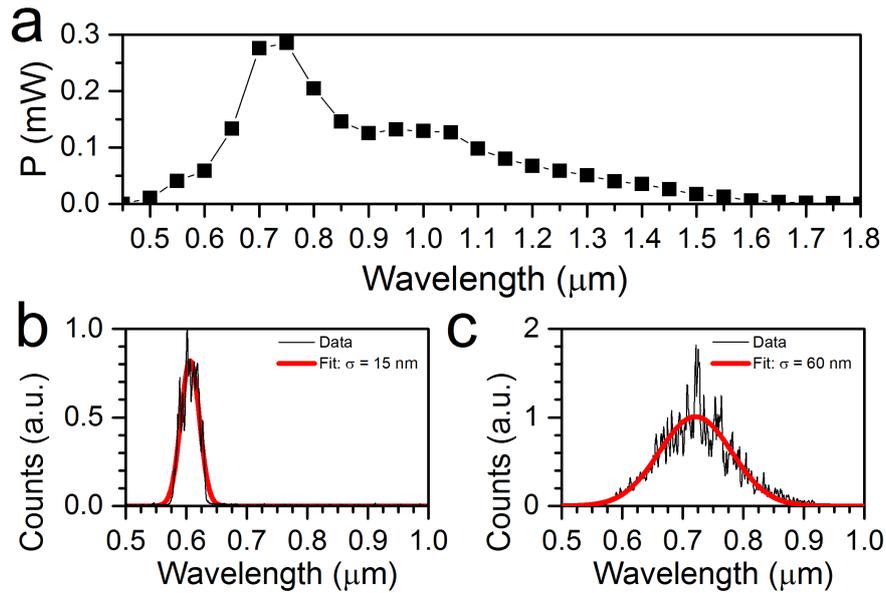

**Figure S1 | Illumination characterization. a,** Measured narrowband spectrum power as a function of center wavelength. **b,** Measured spectrum of a narrowband slice fit by a Gaussian with a standard deviation of 15 nm and a center wavelength of 606 nm. **c,** Measured spectrum of the broader-band slice used to illuminate devices fit by a Gaussian with a standard deviation of 60 nm and a center wavelength of 720 nm.

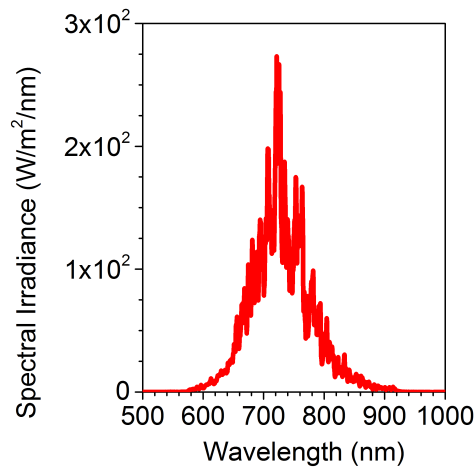

**Figure S2 | Spectral irradiance of experimental spectrum.** Integration of the spectral irradiance results in an irradiance of 17.6 kW/m². Dividing by photon energy and integrating gives a total photon flux of $6.77 \times 10^{22}$ photons/m².

**Shockley-Queisser Detailed Balance Model**

The principle of detailed balance was used by Shockley and Queisser to compute the limiting power conversion efficiency for single-junction pn-junction solar cells [4] (i.e. the Shockley-Queisser limit). To compute the limiting power conversion efficiency, a particle balance model was employed that enabled the computation of current voltage curves from which the figures of merits (i.e. short-circuit current, open-circuit voltage and fill-factor) could be extracted and the power conversion efficiency computed. The particle balance model was physically based upon the fact that all objects emit blackbody radiation and that the radiation emitted by a pn-junction increases exponentially with increasing splitting of the electron and hole quasi-Fermi levels within the pn-junction, that is, with increasing forward bias voltage. This exponentially increasing emission from a pn-junction is caused by the current that flows through the device when it is biased: holes and electrons flow from opposite ends of the pn-junction towards each other and recombine, emitting light. Shockley and Queisser used this radiative recombination current as a physically-based lower-limit to the power-dissipating current which must flow through a pn-junction under bias [4]. Importantly, the Shockley-Queisser limit on the power conversion efficiency of a pn-junction solar cell depends only on the spectrum of the incident light and the bandgap of the absorbing material. Therefore, in our detailed balance modeling, we employ the spectral irradiance of our experimental spectrum (Fig. S2) and the room temperature bandgap of wurtzite InAs, $E_G$ = 0.39 eV.

Three types of particles are tracked in our detailed balance model: 1) absorbed photons, 2) emitted photons and 3) moving charge carriers. The "balancing" of the model is captured in that the absorbed photons, $\Phi_{abs}$ (units of photons per second), must be equal to the sum of the emitted photons, $\Phi_{emt}$ (units of photons per second), and the charge-carriers that move through a connected circuit, $\Phi_{cc}$ (carriers per second):

$$\Phi_{abs} = \Phi_{cc} + \Phi_{emt} \tag{6}$$

Algebraically rearranging, multiplying both sides by the fundamental charge, $e$, (in order to convert moving charge-carriers into current) and using a negative sign to denote a power-producing current, the current voltage curve of the solar cell can be computed as:

$$I(V) = -e\bigl(\Phi_{abs} - \Phi_{emt}(V)\bigr) \tag{7}$$

For the idealized solar cell, we assume that all photons with energy above the bandgap of the absorber are absorbed (and generate one electron-hole pair). This assumption maximizes the photocurrent generation, and consequently, the open-circuit voltage of the cell. Thus, the number of absorbed photons, $\Phi_{\text{abs}}$, is independent of voltage and is computed from

$$\Phi_{\text{abs}} = A \int_{500\text{ nm}}^{\lambda_G} \frac{\lambda F}{hc} d\lambda \qquad (8)$$

where $A$ is the illuminated area, $F$ is the spectral irradiance of our experimental spectrum (Fig. 2a) and we use $E_{\text{photon}} = hc/\lambda$, where $h$ is Planck's constant and $c$ is the speed of light, to convert spectral irradiance into a number of photons per meter-squared per meter of wavelength. The lower integration limit is the shortest wavelength of our source (500 nm, see Fig. S2), and the upper limit is $\lambda_G$ = 3180 nm corresponding to the wurtzite InAs bandgap. Integration of the integrand results in the computation of a photon flux. Then, to convert from a photon flux to a reasonable estimate of the maximum amount of photons absorbed by our nanowire device, we generously assume an absorption efficiency of unity and multiply by the projected surface area of the device. In this case, we consider the device to be the exposed semiconductor nanowire in between the inner contact edges and neglect absorption under the contacts or in portions of the nanowire extending beyond the contact edges. Based upon analysis of SEM and TEM images, the inter-contact length, $L$, is taken to be 400 nm and the diameter, $D$, of a representative nanowire is taken to be 40 nm, giving a projected area of 16×10$^{-15}$ m$^2$.

The voltage-dependent emitted photon number, $\Phi_{\text{emt}}(V)$ - that is, the radiative electron-hole pair recombination current as a function of voltage - is computed as the integral over energy of the modified Planck blackbody radiation equation [5,6] for emission into the full sphere surrounding the device:

$$\Phi_{\text{emt}}(V) = A_{surface} \frac{4\pi}{h^3 c^2} \int_{E_g}^{\infty} \frac{E^2}{\exp\left(\frac{E-qV}{kT}\right) - 1} dE \qquad (9)$$

Note that in Eq. (9) we assume, implicitly, an emissivity of unity for all emission angles corresponding to our assumption of unity of absorption efficiency. To convert from an emitted photon flux to a rate of photons emitted by a cylindrical device, we multiply by our device's surface area: $A_{surface} = \pi D L \approx$ 50×10$^{-15}$ m$^2$. In doing this, we consider the device to be the exposed semiconductor nanowire in between the inner contact edges and neglect emission from under the contacts or from portions of the nanowire extending beyond the contact edges. As in the computation of absorption, the inter-contact length, $L$, is taken to be 400 nm and the diameter, $D$, of a representative nanowire is taken to be 40 nm. When there is no illumination, $\Phi_{\text{abs}}$ is zero, and $\Phi_{\text{emt}}(V)$ can be used to compute the detailed balance limit for dark current through a diode comprised of a material having the bandgap, $E_g$.

Importantly, the above detailed balance limit calculations assume an absorption and emission efficiency of one for our nanowire device across the wavelength range of our experimental spectrum and the device's wavelength range of emission. However, due to diffractive effects in sub-wavelength sized devices, the absorption efficiency and the emission efficiency, $Q_{\text{abs}}$ and $Q_{\text{ems}}$, can in fact be smaller than or larger than 1. We have used electromagnetic modeling to investigate the $Q_{\text{abs}}$ of our nanowire device (Figure S3) and for unpolarized light, we find that $Q_{\text{abs}}$ is less than one within the wavelength range of our spectrum (Figure S3b). Thus, from this modeling, we expect that the absorption performance of the nanowire device leads to a lower short-circuit current density compared to the idealized, unity absorption efficiency case. In the detailed balance analysis, a lower short-circuit current (due to a reduction in $\Phi_{\text{abs}}$ in Eq. (7)) leads to a lower open-circuit voltage. Thus, we rule out absorption enhancement as a possible cause for the high open-circuit voltages observed in our single-barrier devices.

Furthermore, there are two reasons why we can rule out low emission efficiency as the cause of the single-barrier device's high open-circuit voltage. Firstly, we find by inclusion of the emission efficiency into the detailed balance analysis (see below), that an extremely low – and likely, unphysical – emission efficiency is required to achieve a radiatively limited open-circuit voltage as large as that which we experimentally observe. Secondly, we find experimentally that the open-circuit voltage shows extremely strong dependence upon heterostructure resistivity, which would not be the case if geometrically induced emission suppression was the cause of the large open-circuit voltage as this would be the same in structures of different heterostructure, but identical geometry.

To include modification of the emission properties by the nanowire geometry into the detailed balance analysis, we insert $Q_{\text{ems}}$ into Eq. (7):

$$I(V) = -e\big(\Phi_{\text{abs}} - Q_{\text{ems}}\Phi_{\text{emt}}(V)\big) \qquad (10)$$

where $Q_{\text{ems}}$ is the angle and wavelength averaged emission efficiency, which may be greater than or less than the emission efficiency of unity assumed in Eq. (9). In the case $Q_{\text{ems}} < 1$, the solar cell emits, per surface area, fewer photons than a blackbody, which allows for a larger open-circuit voltage at a given $\Phi_{abs}$. When we calculate the dependence of open-circuit voltage on $Q_{ems}$, we find that $Q_{ems}$ = 0.018 is required to reach the experimentally observed $V_{OC}$ = 368 mV (± 5 mV) in the detailed balance analysis (Fig. S4). This calculation assumes that (i) recombination is 100% radiative (that is, there is no non-radiative recombination - which reduces open-circuit voltage from the upper value given by radiative recombination), (ii) $Q_{abs}$ = 1 within the incident spectrum range and (iii) 100% of photogenerated carriers are collected. In the case that any of these three assumptions are not fulfilled, an even larger emission suppression (i.e. lower emission efficiency) is required to reach $V_{OC}$ = 368 mV (± 5 mV). COMSOL wave optic modelling of our experimental system shows that it does not exhibit characteristic (ii) (Figure S3b) and it is extremely unlikely that our unoptimized devices exhibit characteristics (i) or (iii). Therefore, an emission efficiency substantially less than 0.018 is certainly required to reach $V_{OC}$ = 368 mV (± 5 mV). We deem such emission suppression as highly unlikely in our device and therefore, discount it as the explanation for the high open-circuit voltage produced by our single-barrier device. This assessment is strongly supported by our finding that the open-circuit voltage shows extremely strong dependence upon heterostructure resistivity, which would not be the case if geometrically induced emission suppression was the cause of the large open-circuit voltage.

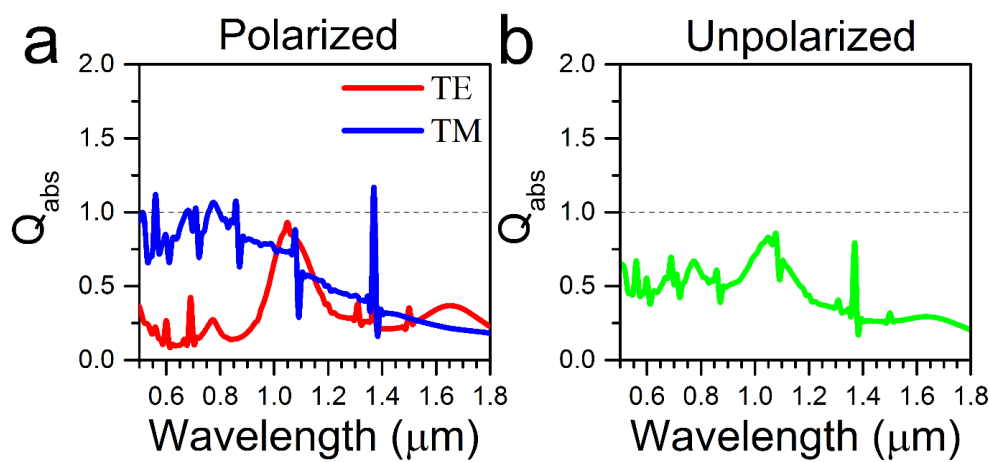

**Figure S3 | COMSOL computed absorption efficiency, $Q_{abs}$, of contacted, single-barrier device. a,** Absorption efficiency for transverse electric (TE) and transverse magnetic (TM) polarizations of light. Transverse refers to the direction of the specified field with respect to the nanowire axis. **b,** Absorption efficiency for unpolarized (i.e. randomly polarized) light. Note that the absorption efficiency $Q_{abs}$ is defined as $n_{abs}/n_{inc}$. Here, $n_{abs}$ is the number of absorbed photons and $n_{inc}$ is the depicted number of photons that would be incident on the

depicted, illuminated geometrical cross-section $A_{geom}$ of the device in a ray-optics description of light. That is, $n_{inc}(\lambda) = A_{geom}I_{inc}(\lambda)/(\hbar c 2\pi/\lambda)$ with $I_{inc}$ the incident intensity. Due to diffractive effects in sub-wavelength sized devices, $Q_{abs}$ can be smaller than or larger than 1.

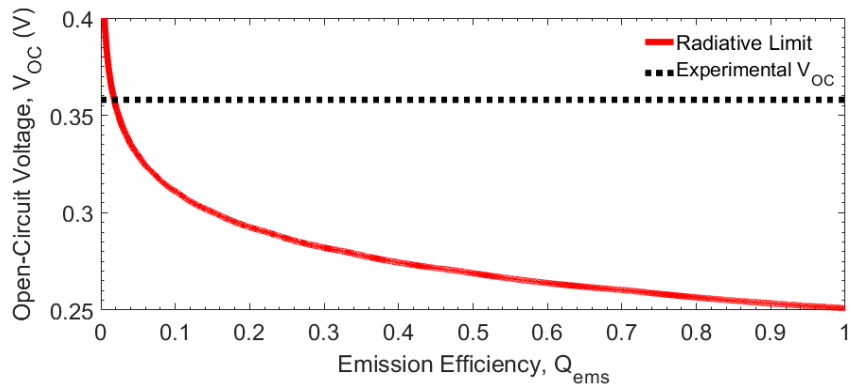

**Figure S4 | Detailed balance limit on open-circuit voltage as a function of emission efficiency.**

**Supporting Information References**: